\PassOptionsToPackage{prologue,svgnames}{xcolor}
\PassOptionsToPackage{numbers, sort&compress}{natbib}
\documentclass[sigconf,screen,nonacm,9pt]{acmart}

\usepackage{subfig}
\usepackage{url}
\usepackage{enumitem}
\makeatletter
\begingroup\endlinechar=-1\relax
       \everyeof{\noexpand}%
       \edef\x{\endgroup\def\noexpand\homepath{%
         \@@input|"kpsewhich --var-value=HOME" }}\x
\makeatother
\def\overleafhome{/tmp}

\usepackage{import}

\ifx\homepath\overleafhome
  \newrobustcmd{\toolspath}[0]{latex-tools-overleaf/}
  \usepackage[frozencache=true,cachedir=.,outputdir=.]{minted}
\else
  \newrobustcmd{\toolspath}[0]{latex-tools/}
  \usepackage[frozencache=true,cachedir=.,outputdir=.]{minted}
\fi

\usepackage{etoolbox}

\newtoggle{AlgsNoEnd}

\newif\iftr     
\newif\ifconf   
\newif\ifnonb   

\usepackage[utf8]{inputenc}
\newcommand{\code}[1]{\texttt{#1}}
\usepackage[binary-units=true]{siunitx}
\usepackage{xspace}
\usepackage[svgnames]{xcolor}

\usepackage{collab}

\usepackage[numbers,sort&compress]{natbib}

\usepackage{soul}
\soulregister\cite7
\soulregister\autoref7
\soulregister\code7
\soulregister\toolname7
\soulregister\ref7
\soulregister\pageref7
\usepackage{hhline}
\definecolor{lightyellow}{RGB}{250, 250, 180}
\definecolor{HLYELLOW}{RGB}{240, 127, 0}
\definecolor{hlyellow}{RGB}{240, 127, 0}
\sethlcolor{lightyellow}

\usepackage{algorithm}
\usepackage{algpseudocode}
\usepackage{pifont}
\usepackage{amsmath}

\algdef{SE}[FOREACHIN]{ForEachIn}{EndForEachIn}[2]{\algorithmicfor\ \textbf{each}\ #1\ \textbf{in}\ #2\ \textbf{do}}{\algorithmicend\ \algorithmicfor}%
\algdef{SE}[FOREACHP]{ForEachP}{EndForEachP}[1]{\algorithmicfor\ \textbf{each}\ #1\ \textbf{in parallel do}}{\algorithmicend\ \algorithmicfor}%
\algdef{SE}[FOREACHINP]{ForEachInP}{EndForEachInP}[2]{\algorithmicfor\ \textbf{each}\ #1\ \textbf{in}\ #2\ \textbf{in parallel do}}{\algorithmicend\ \algorithmicfor}%
\algdef{SE}[FORMOD]{ForMod}{EndForMod}[3]{\algorithmicfor\ #1\ \textbf{from}\ #2\ \textbf{to}\ #3\ \textbf{do}}{\algorithmicend\ \algorithmicfor}%
\algdef{SE}[FORMOD]{ForModS}{EndForModS}[4]{\algorithmicfor\ #1\ \textbf{from}\ #2\ \textbf{to}\ #3\ \textbf{step}\ #4\ \textbf{do}}{\algorithmicend\ \algorithmicfor}%
\algdef{SE}[FORMODP]{ForModPS}{EndForModPS}[3]{\algorithmicfor\ #1\ \textbf{from}\ #2\ \textbf{to}\ #3\ \textbf{in parallel do}}{\algorithmicend\ \algorithmicfor}%
\algdef{SE}[FORMODP]{ForModP}{EndForModP}[4]{\algorithmicfor\ #1\ \textbf{from}\ #2\ \textbf{to}\ #3\ \textbf{step} #4\ \textbf{in parallel do}}{\algorithmicend\ \algorithmicfor}%

\iftoggle{AlgsNoEnd} {
  \algtext*{EndFunction}
  \algtext*{EndIf}
  \algtext*{EndFor}
  \algtext*{EndForEachIn}
  \algtext*{EndForEachP}
  \algtext*{EndForEachInP}
  \algtext*{EndForModP}
  \algtext*{EndForModPS}
  \algtext*{EndForModS}
}{}

\algdef{SE}[VARIABLES]{Variables}{EndVariables}{\algorithmicvariables}
  {\algorithmicend\ \algorithmicvariables}
  \algnewcommand{\algorithmicvariables}{\textbf{global}}
\algnewcommand{\LineComment}[1]{\State \(\triangleright\) #1}
\algnewcommand{\And}{\textbf{and}\xspace}

\usepackage{tikz}
\usepackage{pifont}
\usetikzlibrary{shapes}

\usepackage{listings}
\newfloat{codeblock}{H}{myc}
\definecolor{darkblue}{rgb}{0,0,.6}
\definecolor{darkred}{rgb}{.6,0,0}
\definecolor{darkgreen}{rgb}{0,.5,0}
\definecolor{red}{rgb}{.98,0,0}
\definecolor{gray}{rgb}{.6,.6,.6}
\definecolor{newgreen}{RGB}{169,209,142}
\definecolor{newpurple}{RGB}{237,134,254}
\definecolor{neworange}{RGB}{244,177,131}
\definecolor{newyellow}{RGB}{255,217,102}
\collabAuthor{copik}{blue}{Marcin}
\collabAuthor{lex}{violet}{lex}
\collabAuthor{htor}{magenta}{Torsten}

\usepackage{booktabs}
\usepackage{adjustbox}
\usepackage{multirow}

\usepackage{pifont}
\newcommand{\xmark}{\ding{55}}
\newcommand{\redx}{\textcolor{red}{\xmark{}}}
\definecolor{nicergreen}{RGB}{57,151,92}
\newcommand{\greencross}{\textcolor{nicergreen}{\ding{51}}}
\newcommand{\yellowbar}{\textcolor{orange}{–}}

\newenvironment{bluebox}{%
\noindent

\adjustbox{innerenv={varwidth}[c]{0.9\linewidth},margin=\fboxsep+.25cm \fboxsep+.2cm,bgcolor=blue!10,center}\bgroup
}{%
\egroup
}

\usepackage{wrapfig}

\newcommand{\toolname}{\emph{FMI}\xspace}

\copyrightyear{2023}
\acmYear{2023}
\setcopyright{acmlicensed}\acmConference[ICS '23]{2023 International Conference on Supercomputing}{June 21--23, 2023}{Orlando, FL, USA}
\acmBooktitle{2023 International Conference on Supercomputing (ICS '23), June 21--23, 2023, Orlando, FL, USA}
\acmPrice{15.00}
\acmDOI{10.1145/3577193.3593718}
\acmISBN{979-8-4007-0056-9/23/06}



\begin{document}

\title{FMI: Fast and Cheap Message Passing for Serverless Functions}

\author{Marcin Copik}
\email{marcin.copik@inf.ethz.ch}
\affiliation{%
  \institution{ETH Zürich}
  \country{Switzerland}
}
\authornote{These two authors contributed equally.}

\author{Roman Böhringer}
\email{r.boehringer@opencore.ch}
\affiliation{%
  \institution{OpenCore GmbH}
  \country{Switzerland}
}
\authornotemark[1]

\author{Alexandru Calotoiu}
\email{alexandru.calotoiu@inf.ethz.ch}
\affiliation{%
  \institution{ETH Zürich}
  \country{Switzerland}
}

\author{Torsten Hoefler}
\email{htor@inf.ethz.ch}
\affiliation{%
  \institution{ETH Zürich}
  \country{Switzerland}
}

\begin{abstract}
Serverless functions provide elastic scaling and a fine-grained billing model, making
Function-as-a-Service (FaaS) an attractive programming model.
However, for distributed jobs that benefit from large-scale and dynamic parallelism,
the lack of fast and cheap communication is a major limitation.
Individual functions cannot communicate directly, group operations do not exist,
and users resort to manual implementations of storage-based communication.
This results in communication times multiple orders of magnitude slower than those found in HPC systems.
%
%
We overcome this limitation and present the FaaS Message Interface (FMI).
FMI is an easy-to-use, high-performance framework for general-purpose point-to-point and
collective communication in FaaS applications.
We support different communication channels and offer a model-driven channel selection according
to performance and cost expectations.
We model the interface after MPI and show that message passing can be integrated
into serverless applications with minor changes, providing portable communication closer to that offered by high-performance systems.
In our experiments, FMI can speed up communication for a distributed machine learning FaaS application by up
to 162x, while simultaneously reducing cost by up to 397 times.
\end{abstract}

\maketitle

{\small\noindent\textbf{FMI open-source implementation:} \url{https://github.com/spcl/fmi}}

{\small\noindent\textbf{TCPunch open-source implementation:} \url{https://github.com/spcl/TCPunch}}

\section{Introduction}

%
Function as a Service (FaaS) is an emerging programming paradigm popular
in cloud applications.
In FaaS, users focus on writing application code decomposed into a set of functions.
Users are not concerned with deploying code and managing the underlying
compute and storage infrastructure.
Instead, function invocations are executed by the cloud provider on
dynamically provisioned servers.
Thus, users never allocate servers (\emph{serverless computing}) and are charged
only for computing time and memory resources used (\emph{pay--as--you--go billing}).
Small, stateless functions do not need to communicate -- they simply write their results to storage, and future functions can continue from there.
Thanks to the fine-grained billing model, functions
are a popular programming model for irregular and unbalanced workloads.

%
Functions are used for distributed and stateful computations in data analytics,
linear algebra, processing of multimedia, machine learning, and high-performance computing~\cite{10.1145/3357223.3362711,10.1145/3361525.3361535,Mller2019LambadaID,jiang2021towards,227653,perronStarlingScalableQuery2020,10.5555/3154630.3154660,copik2022software,copik2022softwarepaper}.
These workloads benefit from the fast and cheap scalability of ephemeral function workers.
%
%
However, such functions run longer and have a significant internal state. This makes storing the entire state and continuing execution later once new inputs become available inefficient --- they need a cheap and fast way of exchanging data to become an efficient backend for distributed computations.
Still, they lack a native and high-performance communication interface. 

%
In HPC, communication in a distributed system is done using the Message Passing Interface (MPI). In contrast to virtual machines and HPC applications,
serverless functions execute in sandboxes that provide strict isolation but are prevented from
accepting incoming network connections (Sec.~\ref{sec:background_communication_nat}).
%
%
%
To communicate, functions rely on slow object storage, in-memory caches, and storage optimized for serverless functions ---
these are primarily designed to improve performance~\cite{10.5555/3291168.3291200}, but
introduce a user-managed and persistent component that defeats the purpose of serverless computing.
%
%
Users need a flexible choice between fast and cheap network communication
and slower, more expensive, and durable storage-based exchange. 
%
%
Unfortunately, while serverless computing is an elastic solution for computing
and resource allocations, it remains surprisingly inflexible when it comes to communication.
%
The performance and price of serverless messaging is already a critical problem,
as messages exchanged over object storage come with double-digit millisecond latency
and cost \$6 per million.
%

%
%
The importance of collective operations is known in the HPC community~\cite{https://doi.org/10.1002/cpe.1206,296665}
--- they are used in virtually all MPI jobs~\cite{rabenseifner1999automatic}.
Collectives offer a portable interface for standard parallel programming patterns~\cite{10.5555/2385466}.
Replacing send--receive messages with collectives makes it
easier to program, debug, and maintain, and can boost performance~\cite{10.1145/963778.963780}.
From the user's point of view, collectives provide \emph{"division of labor: the programmer thinks in terms of these primitives and the
library is responsible for implementing them efficiently"}~\cite{10.1016/j.parco.2009.09.001}.
This separation is crucial in the black-box serverless world with major
differences between cloud providers.
At the same time, high performance 
communication requires finely tuned algorithms
according to network topology, number of participants, message size,
application, and even the memory hierarchy~\cite{10.1007/978-3-540-39924-7_38,Tu2012,10.1145/2462902.2462903,10.5555/3291656.3291695,10.1145/3295500.3356222,Pjesivac-Grbovic2007}.
%
%
%
However, the entire communication hierarchy that includes nodes, racks, sockets, processes, and caches
is hidden from the user in serverless.
This is an additional motivation for the cloud provider to implement hierarchical and multi-protocol communication~\cite{1437304},
such that serverless applications benefit from standardized message-passing operations
with high-performance implementations.
The world of collective specializations is rich and remains concealed behind system abstractions,
and serverless should benefit from it (Sec.~\ref{sec:background_collective_communication}).


The community identified the lack of support for efficient group communication as a fundamental
limitation of serverless computing~\cite{schleier2021serverless,DBLP:journals/corr/abs-1902-03383,10.1145/3357223.3362711}:
%
%
%
%
%
applications would benefit from the high performance and versatility of collectives,
but need a framework to hide the complexity of the cloud system.


%
In this work, we provide the first direct general-purpose communication framework for FaaS: the \textbf{FaaS Message Interface (FMI)}.
\toolname{} is an easy--to--use, high-performance framework where the implementation details of point-to-point and collective operations are hidden
behind a standardized interface inspired by MPI,
providing portability between clouds and runtime adaption. We use MPI as our guide as it has established itself as {\em{the}} communication solution for distributed memory systems.
We have implemented and extensively evaluated multiple communication channels with respect to both price and performance, and they are all included in the current library implementation.
Having determined that direct communication over TCP is the best solution in all scenarios,
we also implement a general-purpose TCP hole punching solution to allow functions to communicate directly,
even behind NAT gateways.
%
%
%
While we implement \toolname{} on AWS Lambda, the design of our library is independent of the cloud
provider and can be ported to any serverless provider.
Furthermore, FMI can be wrapped around an existing MPI implementation, allowing for a seamless port of FaaS applications
to HPC clusters.
%
%
%
%
Concretely, we make the following contributions:
\begin{itemize}[topsep=0pt]
  \item We introduce a library for message passing that provides common and standardized abstractions for serverless point-to-point and group communication.
  \item We provide analytical models for communication channels in FaaS and discuss the
    performance-price trade-offs of serverless communication.
  \item We demonstrate the application of \toolname{} to serverless machine learning and present
    a reduction of communication overhead by a factor of up to 162x and reducing cost by up to 397 times compared to existing solutions.
\end{itemize}

\section{Background}
\label{sec:background_faas}
\label{sec:background_communication_mediums}

Distributed FaaS applications already implement many group and collective operations patterns
across concurrent functions, prominent examples being MapReduce in data analytics~\cite{GIMENEZALVENTOSA2019259,Mller2019LambadaID,wawrzoniak2021boxer,10.1007/978-3-642-03770-2_30}, \sloppy \texttt{reduce-scatter} in machine learning~\cite{10.1145/3357223.3362711,jiang2021towards},
and \texttt{scan} in video encoding~\cite{10.5555/3154630.3154660}.
Furthermore, serverless applications need direct communication to offer performance
competetive with persistent servers~\cite{copik2022faaskeeper}.

%
However, the inter--function communication remains the Achilles' heel of serverless. Inspired by the statement: 
\emph{"Storage is not a reasonable replacement for directly-addressed networking, even
with direct I/O —it is at least one order of magnitude too slow."}~\cite{hellerstein2018serverless}, we list different communication channels and conduct a detailed performance (Sec.~\ref{sec:communication_mediums})
and cost analysis of these cloud systems (Sec.~\ref{sec:communication_models}):

%


%
%
%
%

%
%
\begin{itemize}
  \item \textbf{Object Storage.}
    These systems offer persistent storage for large objects with high throughput,
    strong consistency~\cite{10.1145/2043556.2043571}, data reliability~\cite{awsS3,azureStorage,gcpStorage},
    and a cost linear in the number of operations and size of stored data.
  \item \textbf{Key-Value Storage.}
    NoSQL databases offer low latency and throughput scaled to the workload~\cite{awsDB,azureDB}.
    However, they only support small objects (400kB in DynamoDB) and have high costs for write operations.
  \item \textbf{In-Memory and Hybrid Storage.}
    In-memory stores such as Redis~\cite{AmazonElastiCacheRedis} and memcached~\cite{AmazonElastiCacheMemcached}
    offer higher performance at the cost of manual scalability management by
    the user and non-serverless resource provisioning.
    Serverless-optimized storage uses multiple tiers of memory and disk~\cite{10.5555/3291168.3291200}.
    The costs depend on the memory size and the time it remains in use.  

  \item \textbf{Direct Communication.}
    %
    Direct network connections could offer high performance without incurring any costs. We discuss a prototype implementation in Sec.~\ref{sec:collectives}
    %
\end{itemize}

\begin{bluebox}
Today's serverless functions tend to communicate using cloud proxies for messaging.
Functions cannot establish direct connections which would provide higher
performance at a lower cost.
\end{bluebox}

\section{Faas Message Interface}

\label{sec:collectives}
%

In this section, we discuss \textbf{FMI}, the FaaS Message Interface for point--to--point and collective communication as well as the assumptions made by our approach. We then discuss communication channels tested for FMI (Sec.~\ref{sec:channels}).
%
We model the interface of FMI after the proven interface of MPI (Sec.~\ref{sec:collectives_interface}) 
and implement a selection of the most common collective operations in serverless
applications (Sec.~\ref{sec:collectives_collectives}).
We design FMI to be modular - our design  makes no assumptions about the underlying cloud system.
This is crucial as cloud systems change quickly and often contain proprietary components.
FMI can be extended with new communication channels, collective operations,
and support for programming languages other than C/C++ and Python.
%
%

\subsection{Assumptions}
\label{sec:collectives_assumptions}

\textbf{Isolation of serverless functions.}
Small, stateless functions work well in isolation by writing results to storage and can be scheduled independently from each other.
However, the type of functions used in more complex serverless workflows~\cite{10.1145/3357223.3362711,10.1145/3361525.3361535,Mller2019LambadaID,jiang2021towards,227653,perronStarlingScalableQuery2020,10.5555/3154630.3154660}
are neither stateless nor independent of each other, requiring complex task dependencies and communication
over cloud storage.
We dispense with the assumption that FaaS functions should be considered in isolation --- the evolving nature of the serverless computing
makes it deprecated in many practical scenarios.

\textbf{Simultaneous scheduling.}
We assume that all functions that will be part of the same communication entity, or communicator, can be scheduled simultaneously.
A timer is started as soon as the first function joins the group communicator.
If all functions scheduled to join do not do so before the timer expires, then all functions exit with an error. 
%

\textbf{Fault tolerance.}
In FaaS, individual functions can retry on failure.
In FMI, there is no recovery mechanism for individual communicator members.
If a function fails or a communication channel times out, the entire communicator exits with an error.
Users can implement fault-tolerant policies on top of FMI, similar to such approaches in MPI.
We propose using group membership combined with timeouts to ensure all functions are scheduled simultaneously, and that failure is detected.
Furthermore, FMI can be used in combination with techniques such as checkpoint/restart~\cite{4228333}
to provide fault-tolerant execution in the serverless cloud.

\textbf{Direct Communication.}
We assume that direct communication between functions without cloud proxies is possible on the platform. Ideally, cloud providers offer an interface for managed TCP or RMA connection setup.
As an alternative, NAT hole punching is discussed in the next section.
%

\subsection{Communication Channels}
\label{sec:channels}
While communicators are responsible for data conversion and serialization, channels
are the medium for data exchange and operate on raw memory.
We broadly classify them in  \emph{direct} and \emph{mediated} channels.
In mediated channels, the communication is done over storage or other indirect means.

%
Mediated channel examples are object storage  (AWS S3), key-value database (DynamoDB),
in-memory cache (Redis), and we create direct channels using TCP connections.
Programmers could add new channels to the library with little effort and benefit from other existing FMI features - such as the implementations of collective operations.
For example, support for QUIC could be added to provide reliable and secure communication on top of UDP~\cite{10.1145/3098822.3098842}.
If a channel provides more specialized mechanisms, such as support for reductions, it is added by overriding default collective algorithms.

Ideally, cloud providers would provide direct communication between functions as a service.
Until such time, we provide a method that allows direct communication using TCP in the current serverless ecosystem.
We first summarize communication over Network Address Translation (NAT), highlight the obstacles to using it in the serverless ecosystem, and suggest hole punching as a solution.

\subsubsection{Network Address Translation (NAT)}
\label{sec:background_communication_nat}
Function instances are placed in sandboxes behind a NAT gateway~\cite{rfc2663}.
The gateway hides the endpoint by rewriting the internal address with an external one in packet headers.
An outgoing communication creates an entry in the translation table.
This enables replies sent to the external address to be forwarded to the intended recipient.
Packets are dropped when there is no entry in the translation table.
Therefore, the party initiating the communication can be behind a NAT gateway but not the recipient.
Thus, when both parties are behind a NAT gateway, direct communication is not possible.

\begin{figure}[t]
  \centering
  \includegraphics[width=1\linewidth]{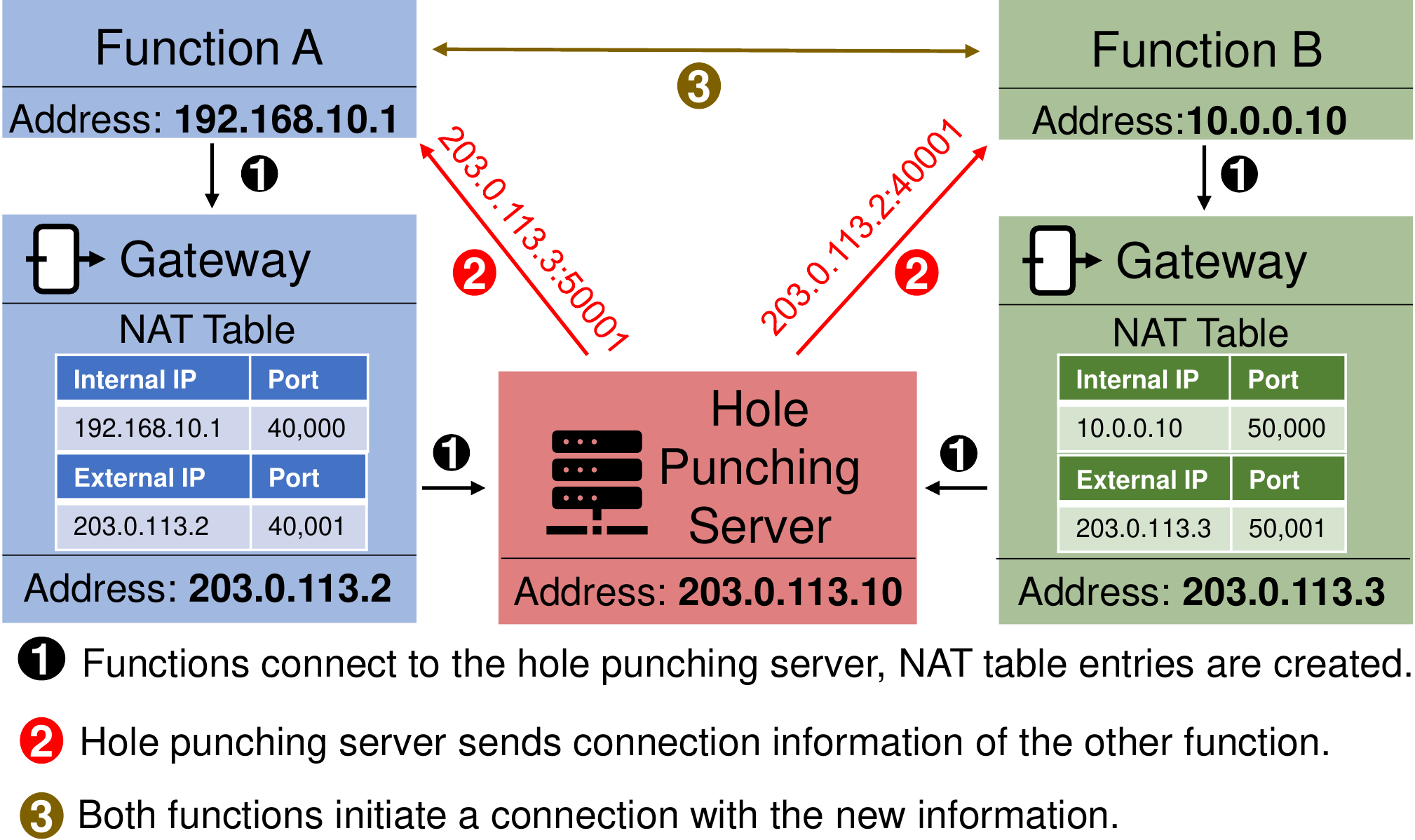}
  \caption{\textbf{Network Address Translation (NAT) Hole Punching.} }
  \label{img:hole_punching_visualization}
\end{figure}

\subsubsection{Hole Punching}
One technique to circumvent the restricted direct communication for endpoints behind a NAT
is hole punching~\cite{fordPeertopeerCommunicationNetwork2005,eppingerTCPConnectionsP2P2005}.
This approach relies on a publicly reachable relay server to create mappings in the translation table
and exchange the other party's address with each participant.
%
%
Functions connect to the hole-punching coordinator and wait for the other party to request a connection.
Then, participants attempt to connect simultaneously using the existing address
mappings from the previous step (Fig.~\ref{img:hole_punching_visualization}).
\subsection{Collective Communication}
\label{sec:background_collective_communication}

\label{sec:collectives_collectives}
Decades of research into collective operations have led to many optimized
communication protocols.
Collective algorithms have different time, memory, and energy trade-offs~\cite{Hoefler_Moor_2014}.
The cost of operations is another fundamental characteristic of communication via cloud storage.
Modern collective operations are extensively tuned: 
MPI collectives are specialized for network transport protocols~\cite{PATARASUK2009117,10.1145/301104.301116,8457871,10.5555/2396095.2396108},
network topology~\cite{9229573,PJESIVACGRBOVIC2007613,1592716,10.1145/2145816.2145823,846009},
and even for specific needs of applications,
such as bandwidth and sparsity optimizations in machine learning~\cite{10.1145/3126908.3126954,10.1145/2966884.2966912,10.1145/3295500.3356222,8514855}.

Cloud providers must be the ones to apply such optimizations
as the abstraction layer prevents users from understanding the system's architecture, and the opportunities for improvement are no less complex than for MPI.
Serverless heterogeneity is increasing with RMA~\cite{copik2021rfaas} as well as GPUs~\cite{satzke2020efficient,kim2018gpu}, and the dynamically
changing topology of workers presents additional challenges~\cite{7530080,6468466,10.1145/3127024.3127031}.



%
We implement the following collective operations from the MPI standard~\cite{mpi-3.0}:
\code{broadcast}, \code{barrier}, \code{gather}, \code{scatter},
\code{reduce}, \sloppy\code{allreduce}, and \code{scan}.
The algorithms selected differ depending on the communication channel used, and should be modified
and updated according to user needs and cloud system configuration.
For example, the \code{scan} operation can use a
depth-optimal but work-inefficient algorithm~\cite{9477174} that is impractical on channels with high data movement cost.
%

\textbf{Mediated channels.}
In \code{broadcast}, the root process uploads the object to the object storage,
and other functions download it, benefiting from the scalable bandwidth of the storage.
In \code{barrier}, each function uploads a 1-byte object and polls until all data is available.
Polling is implemented using the \emph{list} operation on the storage -- this counts the number of objects and succeeds when the count equals
the number of functions in the communicator.
In \code{gather}, functions upload their buffers to storage, and the root
node polls for data while \code{scatter} follows an inverted communication pattern.
Similarly, the root node downloads the data and applies the reduction in \code{reduce} and \code{allreduce}.
Finally, in \code{scan}, each function polls for the partial result of its predecessor, applies the scan operator, and uploads the result.

\textbf{Direct channels.} This case is similar to the MPI use case, so
\code{broadcast}, \code{gather}, \code{scatter}, and \code{reduce} are implemented with a
binomial tree to avoid the bandwidth limitations of a single function.
\code{Allreduce} uses recursive doubling~\cite{10.1007/978-3-540-39924-7_38},
\code{barrier} is implemented as an \code{allreduce} with one-byte input
and the no-op reduction operator and \code{scan} is implemented with a two-phase tree-based operation~\cite{42122,sandersParallelPrefixScan2006}.

\subsection{Implementation}
The FMI library is implemented in roughly 1,900 lines of C++ code,
%
%
as well as \emph{Infrastructure--as--a--Code} in the form of AWS Lambda layers and
CloudFormation templates~\cite{awsCloudFormation}.
%
FMI users can use layers to integrate the message-passing library into serverless applications without
any build steps.
Furthermore, we implement a hole punching library and server, \code{TCPunch}, as we found no open-source solution for C/C++.
%
%
The server stores address translations in memory and disseminate them to parties that are trying to establish connections.
%
%
TCPunch does not contain any FMI-specific logic and exposes a simple interface, 
allowing other applications to easily integrate it\footnote{For details on the source code and configuration of hole punching and message passing,
we refer readers to a technical report: \url{https://www.research-collection.ethz.ch/handle/20.500.11850/532425}}.

\textbf{Portability}
Our approach is platform-agnostic and can be ported to any serverless platform that upholds the main
assumptions (Sec.~\ref{sec:collectives_assumptions}).
New cloud systems can be supported by implementing the interface of mediated channels
in terms of available storage and database systems.
In particular, FMI can be used in open-source, self-hosted, and Kubernetes-based serverless platforms without
any modifications.
FMI requires system administrators only to deploy hole punching instances on virtual machines and containers.

Furthermore, FMI can be seamlessly ported to HPC systems by wrapping the MPI library to benefit from optimizations and
support for high-speed networking that existing MPI implementations provide.
Thus, a single codebase can be used for scaling parallel processing on both MPI ranks in an HPC cluster and serverless
functions in a public cloud.

\textbf{Practical Experiences}
We have deployed FMI with NAT hole punching successfully on the AWS cloud.
However, we observed unexpected timeouts between some functions, which required repeating the hole-punching procedure to create a new pair of connected sockets.
An analysis reveals that the source of the problem is TCP acknowledgments that never arrive at the destination, caused by the interference of NAT gateways with TCP timestamps and using Ethernet jumbo frames on EC2 virtual machines.
Timestamps are enabled by default in the TCP stack of many Linux distributions.
Disabling them resolves the issue in user-controlled environments like virtual machines, self-hosted serverless platforms, and Kubernetes instances.
However, preventing the issue entirely in environments such as AWS Lambda requires support from the cloud provider, as functions running on this platform operate in a restricted environment.


\subsection{Interface}
\label{sec:collectives_interface}
The FMI interface is based on MPI, so programmers familiar with MPI can use it without adjustment.
The interface is designed with compatibility in mind: we primarily
extend the MPI interface with modern C++ features, e.g., we remove the need for explicit typing
in many operations: 
\begin{minted}[fontsize=\footnotesize,frame=lines,]{cpp}
#include <fmi.h>
// The functions are part of a communicator comm
// Here, the communicator contains 3 functions
// Each function has a unique id: 0, 1, 2
// Defining send buffer:
FMI::Comm::Data<std::vector<int>> vec({0, 1, 2});
// Defining receive buffer:
FMI::Comm::Data<std::vector<int>> recv(1);
// Collective operation
comm.scatter(vec, recv, 0);
// Test that each function got the correct data
assert(recv.get()[0] == my_id);
\end{minted}

\textbf{Languages.}
Support for new languages can be easily added by implementing a wrapper around
the communicator library.
We demonstrate the support for Python, with the help of the \texttt{Boost.Python} library:
\begin{minted}[fontsize=\footnotesize,frame=lines]{python}
import fmi
# Defining a datatype:
dtype = fmi.types(fmi.datatypes.int)
# Root function sends data:
if my_id == 0:
    comm.bcast(42, 0, dtype)
# All other functions receive data:    
else:
    assert comm.bcast(None, 0, dtype) == 42
\end{minted}

\begin{table*}[t]
\centering
  \begin{adjustbox}{max width=\linewidth}
  \begin{tabular}{lcccccccc}
    \toprule
    Channel & Latency & Bandwidth & Cost    & Scalability   & Max. Message &  Push vs Pull? & Message Persistence & Serverless? \\
    \midrule
    Object Storage  & Very High  & Low  & Low   & Provider-side     & 5 TB     & Pull  & \greencross & \greencross \\
    NoSQL Database  & High   & Very Low   & High  & Provider-side     & 400 kB          & Pull  & \greencross & \greencross \\
    In-Memory Cache & Low   & High  & Low   & User-side        & 512 MB   & Pull  & \yellowbar  & \redx \\
    Direct TCP      & Very Low   & High  & Free  & User-side       & Unlimited  & Push  & \redx       & \greencross \\
    \bottomrule
  \end{tabular}
  \end{adjustbox}
\caption{Serverless communication channels. In the following sections, we quantify the differences and characterizations.}
\label{tab:channels_summary}
\end{table*}

\textbf{Communicators.} As in MPI, all message-passing operations are based on the concept of a communicator.
Each communicator is uniquely named and is based on a \emph{group} of $N$ FaaS functions, each one with
a unique identifier in the range $[0, N)$~\cite{5362477}.
Since each function pair in a communicator is independent of each other, communicators can parallelize
the hole punching process to accommodate the increasing number of functions.
Therefore, an application can create multiple communicators with different numbers of peers or lifetimes, and providing the flexibility needed to support the many communication patterns of serverless.
%
%
For collectives that reduce data, such as \texttt{(all)reduce} and \texttt{scan},
users can provide an arbitrary function object as a reduction operation.

Concurrent invocation of multiple parallel functions can be implemented in existing FaaS systems, and
we envision that future serverless runtimes will provide this natively.
%
%

\section{Communication Channel Performance }

\label{sec:communication_mediums}

An ideal communication channel should support both \textbf{low latency}
and \textbf{high throughput} communication.
Many cloud technologies can be used for serverless communication, but none
fulfills all requirements (Tab.~\ref{tab:channels_summary}).

Additional requirements for serverless storage include \textbf{elastic scaling} with serverless
parallelism and efficient support for \textbf{arbitrary object sizes}~\cite{216007}.
The latter is necessary to support the different communication patterns of serverless applications
that can involve both fine-grained messaging and exchanging large data objects.
Furthermore, serverless storage offers message persistence for fault tolerance.
In-memory caches can access some past messages while the instance is running but are lost upon releasing the resource.
While persistence can be useful for fault tolerance, not all application require saving each message.

An aspect to consider is that not all communication channels support \emph{push} messages --- messages where the receiver
blocks and waits for data to arrive.
Instead, the recipient must actively and frequently poll the channel to
verify if the expected message is available --- also known as \emph{pull} messages.
Active polling introduces additional complexity (Sec.~\ref{sec:communication_mediums_benchmarking})
and adds an important performance-cost trade-off (Sec.~\ref{sec:communication_models}).
%

%

Finally, not all systems support a truly serverless deployment where no resource provisioning by the user is required.
Until establishing direct communication channels is offered as a service by cloud providers a hole-punching server must be set up -- it requires minimal resources as its only responsibility is to accept connection requests.
Many tenants can use hole punching simultaneously, and the service scales horizontally according to the traffic.
%

On the other hand, setting up the Redis cluster requires a significant amount of work,
and right-sizing the cluster is the user's responsibility.
The system traffic must be monitored since an underprovisioned Redis cluster will not lead to
failures but instead cause performance degradation, complicating server management further.

To understand the performance implications of selected cloud communication channels,
we consider two scenarios:
With a single sender and a single receiver, we examine \emph{point--to--point} communication, the basic building block of all communication operations (Sec.~\ref{sec:communication_mediums_p2p}).
Then, we consider a \emph{one-to-many} scenario with one sender and a variable number of receivers.
This benchmark intentionally 
stresses bandwidth scalability of the different channels (Sec.~\ref{sec:communication_mediums_group}), so we explicitly do not use algorithmical optimizations in this step.

\begin{figure*}[!th]
  \centering
  \subfloat[One byte message.]{
    \includegraphics[width=0.50\textwidth,page=1,clip,trim={0cm 7.5cm 0cm 0cm}]{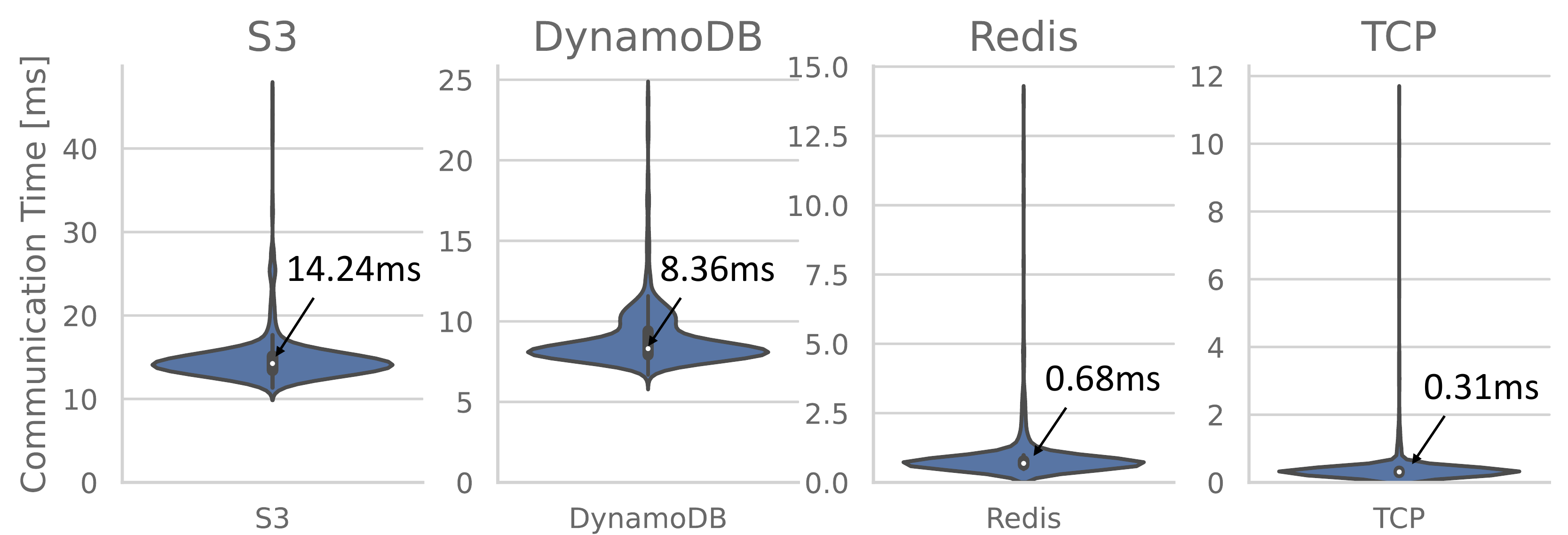}
    \label{fig:p2p_experiment_small}
  }
  \subfloat[1 MB message.]{
    \includegraphics[width=0.43\textwidth,page=1,clip,trim={0cm 4cm 0cm 0cm}]{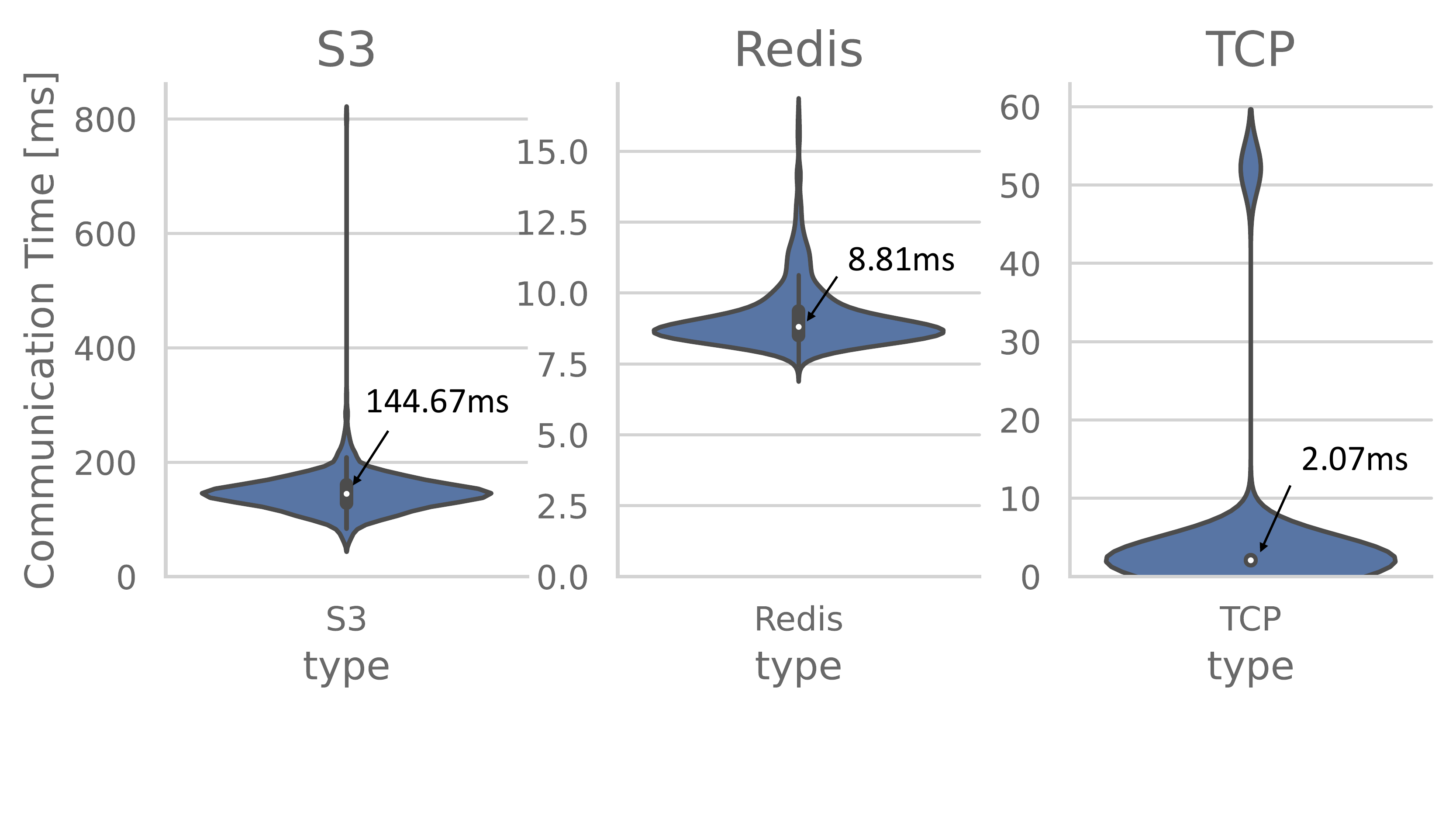}
    \label{fig:p2p_experiment_large}
  }
  \caption{Point--to--point communication latency for messages with one byte (left) and 1 MB (right), 1000 repetitions.}
  \label{fig:p2p_experiment}
\end{figure*}

\begin{figure}[!thb]
  \centering
    \includegraphics[width=1\linewidth]{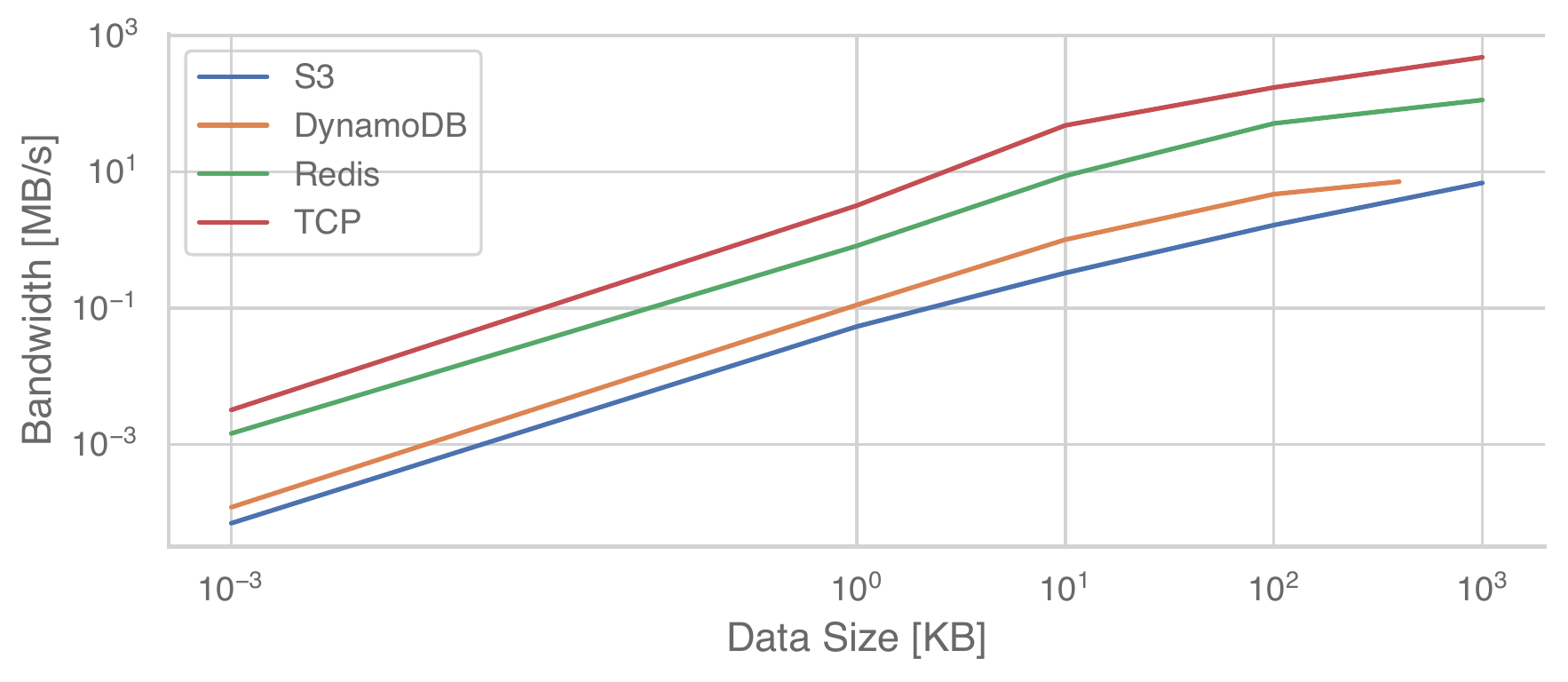}
  \caption{Bandwidth of point--to--point communication with varying sizes.}
    \label{fig:p2p_experiment_bandwidth}
  
  \end{figure}

\subsection{Benchmarking Setup}
\label{sec:communication_mediums_benchmarking}
We analyze the following communication channels in the AWS cloud:
S3 (object storage),
ElastiCache Redis (in-memory data store),
DynamoDB (NoSQL key-value store),
and direct TCP communication with NAT hole punching.
For all of them, we implement the message exchange in serverless Lambda functions written in C++.
We assign 2 GiB of RAM to Lambda functions to provide functions with sufficient I/O bandwidth and
decrease the likelihood of functions' co-location in a single virtual machine~\cite{246184}.
We run the experiments in the cloud region \texttt{eu-central-1}.

The S3 and DynamoDB stores do not require additional configuration beyond creating cloud resources.
We use the pay--as--you--go billing model for DynamoDB, and we deploy Redis on the
\texttt{cache.t3.small} instance with 1.37 GiB RAM and two vCPUs.
The hole punching server fits on a \texttt{t2.micro} instance (1 GiB RAM, one vCPU) as it only stores a few bytes per
connection during setup and it costs less than 1.5¢ per hour.

\emph{Polling.}
To communicate over S3, DynamoDB, or Redis, the producer creates an object or an item in a predetermined
location.
Unfortunately, there is no explicit notification mechanism to inform consumers that data is available.
It is possible to launch new functions asynchronously on data updates but not notify existing functions.
For small, short-running functions, this is no issue.
However, for large stateful functions stopping and resuming is inefficient.
Therefore, consumers must poll the store using the predetermined key until they get the data.
We implement a hybrid backoff strategy to reduce the number of required GET operations,
as each one million reads costs approximately \$0.5: for the first 100 retries, the backoff time is linearly increased from 1 ms to 100 ms, followed by setting the backoff time to 2x retries, i.e., 202 ms, 204 ms. We bound the maximum number of retries to 500.
This is unnecessary for ElastiCache Redis as polling here does not incur additional costs - in
contrast to cloud-managed S3 and DynamoDB, read requests in Redis are free.

\subsection{Point--to--Point}
\label{sec:communication_mediums_p2p}

To measure point--to--point communication, we execute a ping-pong benchmark
and report half of the round-trip time.
For storage-based communication, the time includes both put and get requests.
%
%
For small messages (Fig.~\ref{fig:p2p_experiment_small}), inter-function TCP is the fastest and can achieve microsecond latency.
For large messages (Fig.~\ref{fig:p2p_experiment_large}), direct communication over TCP remains
the fastest option with a reasonably symmetrical density, concentrated around the mean.
%
%
%

For the next experiment, we vary message size from 1 byte to 1 MiB and present the median
bandwidth (Fig.~\ref{fig:p2p_experiment_bandwidth}).
Direct communication remains the fastest communication channel for all data sizes, with the
difference to cloud storage being smaller for larger sizes: the relative overhead of the cloud proxy becomes smaller compared to the transmission time.
While other authors have reported high read bandwidths for S3~\cite{216007}, our measurements include
both the send--receive (put--get) communication and the overhead of polling.
The write bandwidth of S3 cannot currently exceed 70 MB/s for objects up to 10 MB~\cite{216007}, limiting throughput.

\begin{figure}[!thb]
  \centering
 
    \includegraphics[width=1\linewidth]{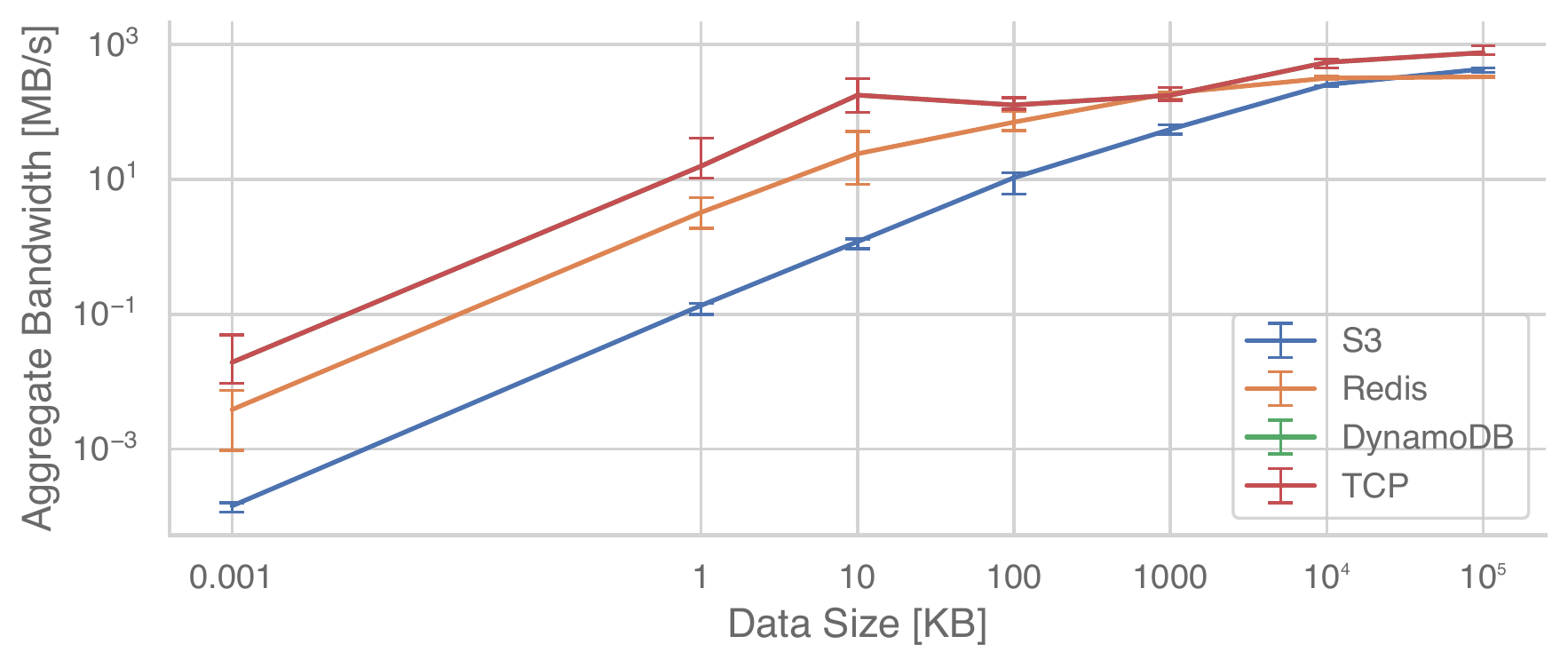}
  \caption{Bandwidth of one-to-many communication with varying message size and 8 receivers.}
    \label{fig:bcast_bandwidth_consumers}
 
\end{figure}

\subsection{One--To--Many}
\label{sec:communication_mediums_group}




To test a one-to-many communication, we use one Lambda producer that sends messages to multiple consumers.
This experiment allows us to assess bandwidth limitations and performance with
multiple functions receiving from one source. 
%
%
%
%
We vary the message size and present the aggregated bandwidth
across all 8 receivers in Fig.~\ref{fig:bcast_bandwidth_consumers}.
While direct communication results in the highest bandwidth, the difference to Redis and S3 decreases
with the number of participants.
%
%
For messages larger than 10 KiB, there is a strong increase in communication time
due to bandwidth limitations.
Both Redis and S3 scale well in this benchmark, as the latter offers automatic
scaling with user count.

We investigated the feasibility of increasing the number of consumers beyond 64.
S3 handles scalability with 128 consumers well, but we observe irregular failures
on Redis with 128 and 256 consumers, likely due to resource limitations.
Our hole punching server easily supported the connection setup with 256 functions, even with a
\texttt{t2.micro} instance.
However, the producer's bandwidth becomes the bottleneck as the number of
consumers grow, highlighting the need for specialized algorithms for collective communication.
%

  %

\section{The Price of Performance}

\label{sec:communication_models}

When modeling and designing HPC collectives~\cite{,Tu2012,10.1145/3284358}, time, memory, and energy trade-offs~\cite{Hoefler_Moor_2014} must be considered. In the serverless world, we must also consider the price of cloud
operations.
The price--performance trade-off has always been a major issue in serverless~\cite{copik2021sebs}:
allocating more powerful instances decreases computation time and resource occupancy but does
not always lead to lower costs.
Therefore we include both the cost of data transfer and the runtime functions spent transmitting data.
We use the alpha-beta model for time --- one of the simplest ways to describe parallel communication.
This model considers $\alpha$, the latency of the communication channel, and $\beta$, the inverse of its bandwidth.
The time to send a message of size $s$ becomes $T=\alpha+s\cdot \beta$.

To compare direct and mediated channels we consider the latency and bandwidth of two functions in a point-to-point communication. We report parameter values for AWS in Tab.~\ref{tab:awsperfmodelparams}. The results show that the in-memory store outperforms object storage in both bandwidth and latency, but they are both inferior to direct communication.
\begin{table}
  \small

    \centering
    \begin{tabular}{lclc}
      \toprule
      Bandwidth & Value & Latency & Value \\
      \midrule
      $\mathtt{1/\beta}(\mathtt{s3})$ & 50 MB/s & $\alpha(\mathtt{s3})$ & 14.7 ms\\
      $\mathtt{1/\beta}(\mathtt{ddb})$ & 7 MB/s & $\alpha(\mathtt{ddb})$ & 8.9 ms\\
      $\mathtt{1/\beta}(\mathtt{redis)}$  & 100 MB/s & $\alpha(\mathtt{redis})$ & 0.88 ms\\
      $\mathtt{1/\beta}(\mathtt{direct})$ & 400 MB/s & $\alpha(\mathtt{direct})$ & 0.39 ms\\ 
      \bottomrule
    \end{tabular}
    \caption{Performance model parameter values for AWS (S3, DynamoDB, ElastiCache Redis, and Lambda).}
    \label{tab:awsperfmodelparams}
\end{table}

We consider the cost per second of executing serverless functions, hosting in-memory cache,
and using cloud storage (Tab.~\ref{tab:aws_cost_model}). We do not incorporate the fixed fee per function invocation in this analysis, as these costs are the same for each communication channel and are negligible for long-running functions.

\textbf{Direct communication.}
As we have seen in Sec.~\ref{sec:communication_mediums_group}, TCP communication has
no inherent cost, but limited bandwidth -- and thus the increased communication time might generate higher costs.
The cost of direct communication is limited to the runtime spent in communication by participating FaaS functions, but we consider the hole punching service currently needed by adding the cost of the virtual machine ($p_\mathtt{hps}$) running the service.

\textbf{Mediated channels.}
For object storage or in-memory data stores, we define a minimal transfer time as the least time it takes for a piece of data to be written to the intermediary storage by a function and then read by another - therefore, assuming there is no time spent waiting and polling for the data to become available. 
The actual transfer time will be longer in practice, as the functions sending and receiving are unlikely to be perfectly synchronized. This will require additional polling on the part of the receiving function, and therefore result in delays.
When using object storage such as S3 or DynamoDB for communication, there are no additional infrastructure costs because the system is managed by the provider and charged on a per-use basis. We further assume that all data is ephemeral and immediately deleted after execution, which leads to negligible storage costs. We therefore only pay for the uploads ($p_\mathtt{s3,u}$, $p_\mathtt{ddb,u}$) and downloads ($p_\mathtt{s3,d}$, $p_\mathtt{ddb,d}$). For communication over an in-memory data store, only infrastructure costs for the store are incurred ($p_\mathtt{redis}$), and these only depend on how long the instance runs.

\begin{table}
      \small
      \centering
      \begin{tabular}{lll}
        \toprule
        Item & Value (\$) & Description\\
        \midrule
        $p_\mathtt{faas}$ & $1.67 \cdot 10^{-5}$ & Lambda GiB per s.\\
        $p_\mathtt{hps}$   & $3.72 \cdot 10^{-6}$    & \texttt{t2.micro} EC2 instance per s. \\
        $p_\mathtt{redis}$  & $1.05 \cdot 10^{-5}$     & \texttt{cache.t3.small} ElastiCache per s. \\
        $p_\mathtt{s3,d}$    & $4.3 \cdot 10^{-7}$     & S3 GET per request. \\
        $p_\mathtt{s3,u}$    & $5.4 \cdot 10^{-6}$     & S3 PUT per request. \\
        $p_\mathtt{ddb,d}$    & $7.62 \cdot 10^{-8}$     & DynamoDB read per 1kB. \\
        $p_\mathtt{ddb,u}$    & $1.5 \cdot 10^{-6}$     & DynamoDB write per 1kB. \\
        
        \bottomrule
      \end{tabular}
      \caption{Price components of the model for AWS in \texttt{eu-central-1}, US dollars.}
      \label{tab:aws_cost_model}
\end{table}

%

%
%
%

\textbf{Cost of FaaS functions.}
The total cost of communication is the sum of the cost to run the FaaS functions during the communication and the cost of moving the data through the channel.
One exchange with $P$ participants, each with $M$ GiB of RAM, that takes $t$ seconds on average has a cost that can be calculated as follows:
\begin{equation}
c_{\mathtt{function}} = P * t *  p_\mathtt{faas} * M
\end{equation}
The cost of Lambda instances increases linearly with memory allocation, hence the $M$ term in the cost equation. The average used for the time $t$ may not necessarily be representative for the individual experienced communication times, but remains a useful approximation given the large number of FaaS functions serverless systems commonly handle.

\begin{table}
   \small

    \centering
    \begin{tabular}{lrrrr}
      \toprule
      Channel & Time (ms) & FaaS (\$) & Channel (\$) & Total (\$) \\
      \midrule
      S3 & 16.70 & $1.12 $ & $5.83 $ & $6.95$ \\
      DynamoDB &151.76&10.10&1,580.00&1,590.10\\
      Redis & 10.88& $0.73 $ & $0.16 $ & $0.84 $ \\
      Direct & 2.89& $0.19 $ & $0.01$ & $0.20$ \\
      \bottomrule
    \end{tabular}
    \caption{Price analysis for communication over S3, DynamoDB, ElastiCache Redis, direct TCP communication.}
    \label{tab:prices}
\end{table}

\textbf{Price-performance analysis.} We  compute the cost and time required by each communication channel to communicate 1MB between two 2GiB Lambda functions a million times by instantiating the models previously described and present the results in Tab.~\ref{tab:prices}. Direct communication is more than four times cheaper \emph{and} faster than all alternatives. 

\section{Evaluation}

\label{sec:evaluation}
We now evaluate the performance and efficiency of our message-passing interface. We focus our evaluation on collective operations, as the point-to-point performance has been analyzed in Sec.~\ref{sec:communication_mediums_p2p}. 
We first examine the scaling of FMI Collectives on AWS Lambda (Sec.~\ref{sec:evaluation_fmi_collectives}).
We then compare FMI's performance against MPI on virtual machines (Sec.~\ref{sec:evaluation_mpi_benchmark}),
and quantify the overheads brought by the serverless environment (Sec.~\ref{sec:evaluation_mpi_faas}).
Finally, we demonstrate how the integration of FMI into a serverless machine
learning framework improves performance and decreases costs (Sec.~\ref{sec:lambdamlbenchmark}).

\begin{figure}[h]
    \centering
    \includegraphics[width=1\linewidth]{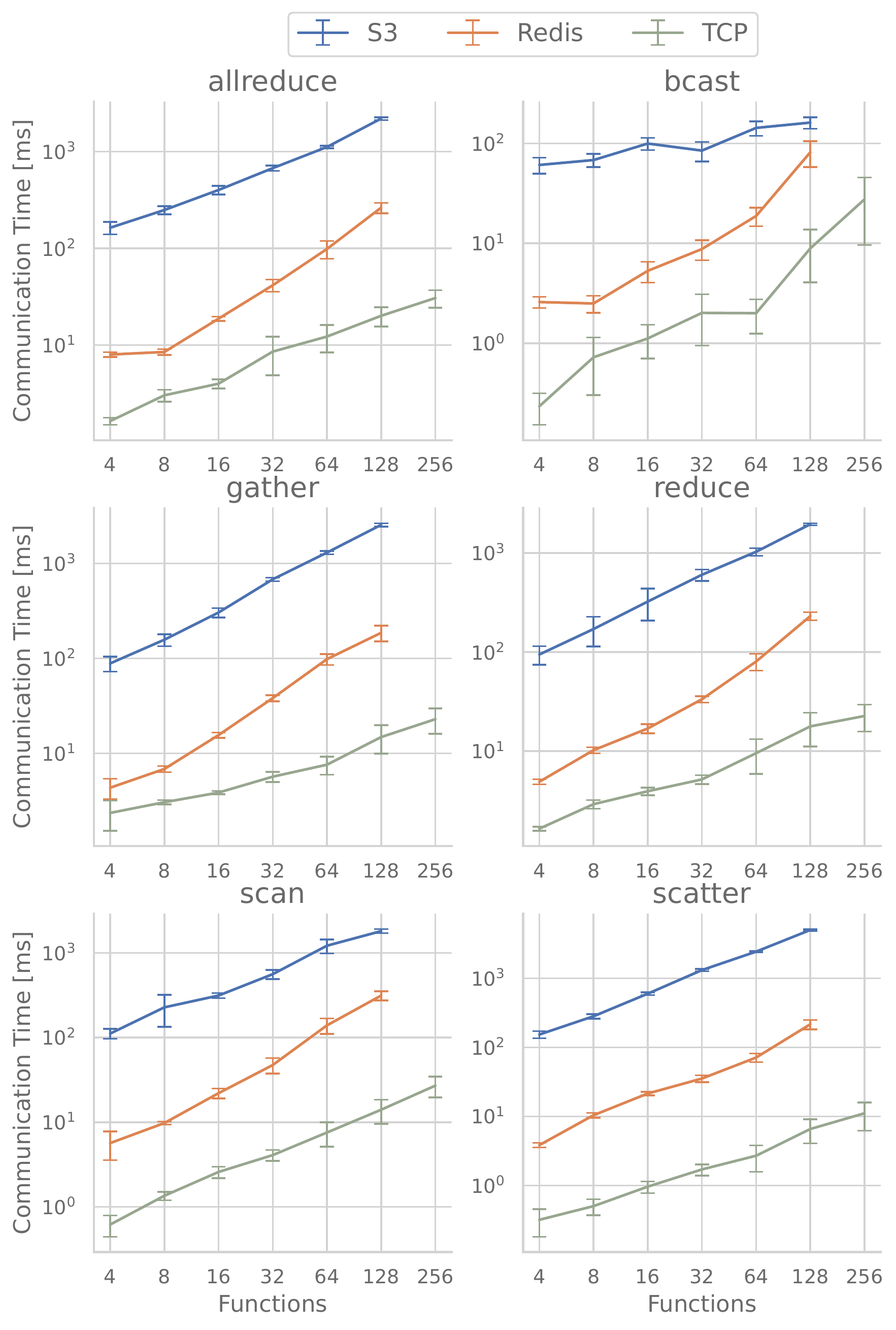}
    \caption{Evaluation of FMI collectives on AWS Lambda.}
    \label{img:smi_benchmark_collectives}
\end{figure}

\subsection{Performance of FMI Collectives in FaaS}
\label{sec:evaluation_fmi_collectives}
To evaluate FMI's collectives on a FaaS platform, we use AWS Lambda functions with 2 GiB RAM.
We deploy C++ functions using the native runtime for AWS Lambda and compile functions with GCC 9.5.
We use AWS SDK C++ 1.9.225 for communication over Redis and AWS, and native TCP/IP sockets.
For Redis, we use one \texttt{cache.t3.small} (1.37 GiB RAM, 2 vCPUs) instance,
and set the polling interval for S3 to 20ms.

We evaluate collective operators with the following operations:
\begin{itemize}
    \item{\textbf{allreduce}}: Adding an integer per process.
    \item{\textbf{bcast}}: Broadcasting an integer.
    \item{\textbf{gather}}: The root receives 5,000 integers in total.
    \item{\textbf{reduce}}: Adding an integer per process.
    \item{\textbf{scan}}: Prefix sum with an integer per process.
    \item{\textbf{scatter}}: The root sends 5,000 integers in total.
\end{itemize}
Each operation is preceded with a \textbf{barrier} to synchronize workers, and we measure the maximum time across all workers needed to complete the collective operation.
We repeat all experiments 30 times.

The results are summarized in Figure~\ref{img:smi_benchmark_collectives}.
In Redis, communication times grow sharply for some algorithms.
This demonstrates the limitation of using a provisioned service where the user
is responsible for scaling resources, and the correct instance size is not 
always known a priori.
Choosing the minimal size that supports a given workload can be
time-consuming and expensive as communication times gradually increase,
which often leads to overprovisioning.
S3 storage performs the worst on all benchmarks due to high latency on small objects,
and only a broadcast operation with a large number of workers demonstrates the benefits of automatic
bandwidth scalability.
On all algorithms, the direct TCP communication achieves the lowest latency.
%

Furthermore, we observe that the average memory consumption does not exceed 100 MB.
However, functions need large memory allocations to achieve sufficient network bandwidth as resources
are scaled with the memory alloction~\cite{copik2021sebs}.

%
%

\begin{bluebox}
  Direct TCP communication enabled by FMI is necessary to achieve high
  performance collective operations.
\end{bluebox}

\begin{figure}[h]
    \centering
    \includegraphics[width=1\linewidth]{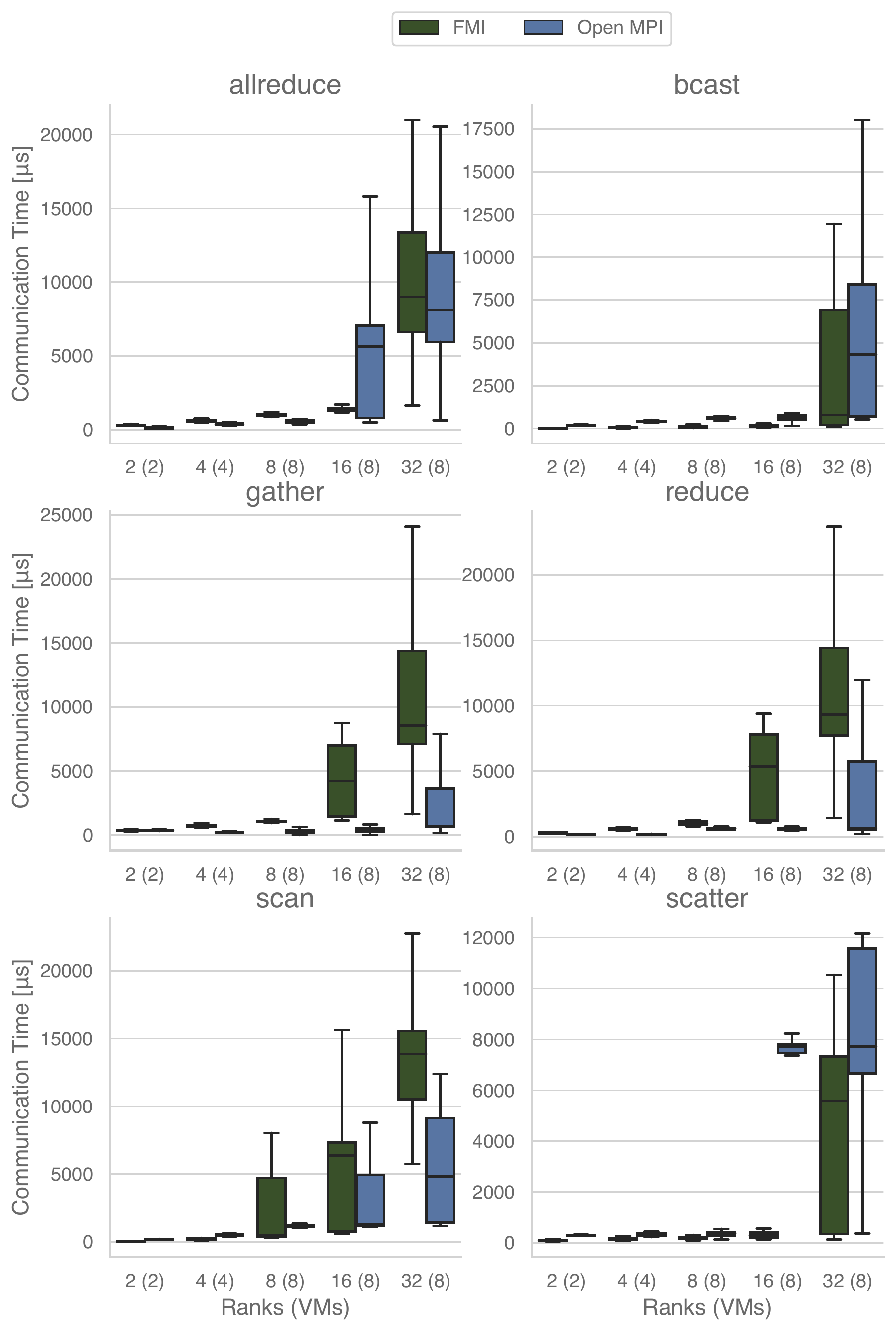}
    \caption{Comparison of FMI with Open MPI in virtual machines. Functions and MPI ranks are spread across virtual machines with 4 vCPUs each.}
    \label{img:smi_benchmark_mpi}
\end{figure}



\subsection{Comparison of FMI and MPI in Virtual Machines}
\label{sec:evaluation_mpi_benchmark}
To compare the performance of FMI and MPI, we deploy both on
virtual machines for an unbiased comparison, 
since MPI is not available on FaaS platforms.

\emph{Setup.}
We execute the MPI benchmarks on \code{t2.xlarge} virtual machines, running
Ubuntu 20.04.1 VMs with 16 GiB of RAM and 4 vCPUs.
%
%
We configure both Open MPI and FI to use one rank per node when using up to eight peers,
and we use up to 4 processes per node otherwise.
We use Open MPI 4.0.3 with the default configuration, which 
uses the TCP transfer layer for communication.
We use a non-tuned FMI installation with direct communication.
%
%

\emph{Performance.}
The performance and variance of FMI collectives is comparable to Open MPI (Fig.~\ref{img:smi_benchmark_mpi}).
We repeat the evaluation of each collective operation 1,000 times, after discarding a first warm-up measurement,
and use a barrier before each experiment.
%
Our implementation of the collectives is competitive and our framework does
not introduce significant overhead.

\begin{bluebox}
FMI is competitive with established MPI implementations, bringing the HPC message-passing performance closer to the serverless world. 
\end{bluebox}

\begin{figure}[h]
    \centering
    \includegraphics[width=1\linewidth]{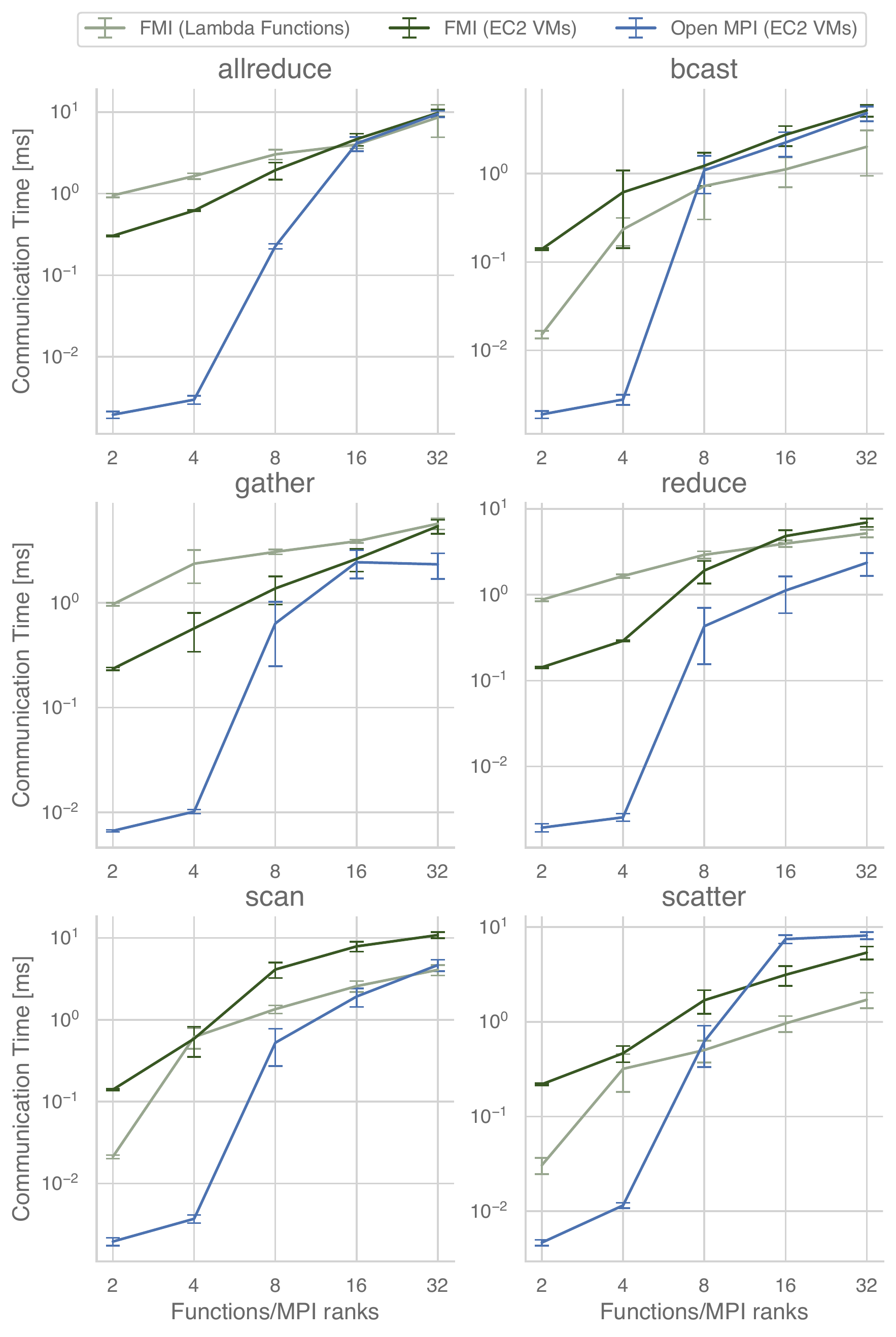}
    \caption{Comparison of collectives on AWS EC2 and AWS Lambda.}
    \label{img:smi_mpi_lambda_benchmark_comparison}
\end{figure}

\subsection{Evaluating the Overhead of FaaS Platforms}
\label{sec:evaluation_mpi_faas}
Thanks to FMI, we can quantify the performance losses incurred by the
serverless environment by comparing operations executed in virtual machines (IaaS) and FaaS.
We deploy FMI and OpenMPI again on virtual machines and use the default co-location settings of OpenMPI.
Then, we compare it against FMI deployment in serverless functions (Sec.~\ref{sec:evaluation_fmi_collectives}).
Results presented in Figure~\ref{img:smi_mpi_lambda_benchmark_comparison} show that communication performance on serverless functions can be worse, even if using identical software and communication algorithms.
Furthermore, the better performance achieved by MPI is explained by its ability to use shared memory for communication when ranks are located on the same machine.
On the other hand, the opaque infrastructure in serverless functions prevents using local means of communication, as functions are unaware of co-location and always use the NAT hole punching for communication, even when both parties are on the same machine.

\begin{figure}[h]
    \centering
    \includegraphics[width=1\linewidth,page=1,clip,trim={0cm 0cm 4cm 0cm}]{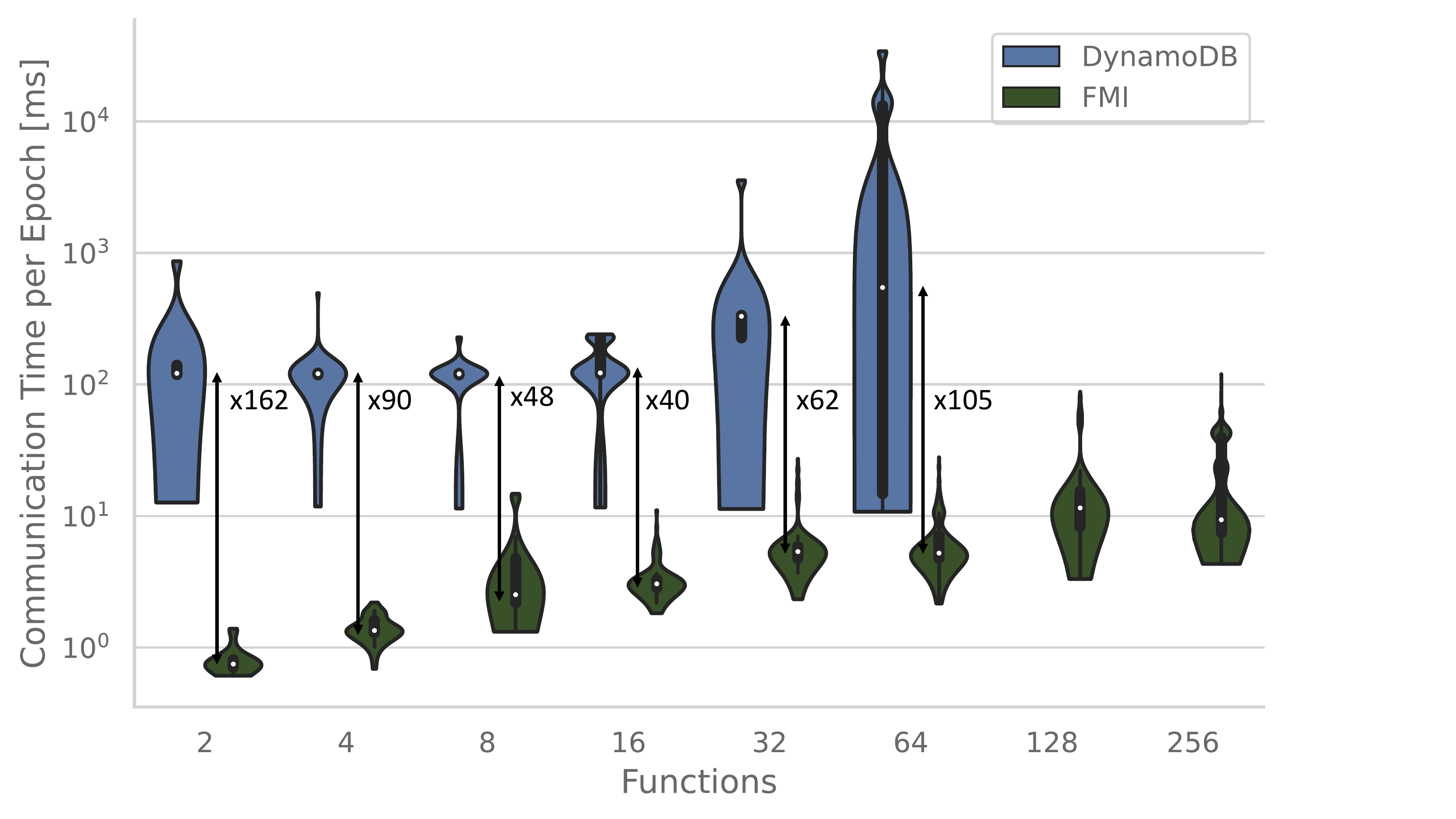}
    \caption[]{Performance comparison of FMI and the LambdaML DynamoDB communication time.\footnotemark{}}
    \label{img:smi_benchmark_lambdaml}
\end{figure}

\footnotetext{The annotation denotes the speedup of the median communications time provided by FMI. No measurements beyond 64 functions are provided for DynamoDB due to outliers and timeouts.}







\subsection{Practical Case Study: Distributed Machine Learning}
\label{sec:lambdamlbenchmark}
To demonstrate the benefits of integrating FMI into serverless applications,
we use LambdaML, a state-of-the-art framework for distributed machine learning
on AWS Lambda, using 
distributed K-Means with the DynamoDB
backend as it was shown to be the best performing~\cite{jiang2021towards}.
%
For FMI, we replace the allreduce provided by the author with
the corresponding FMI collective.
We use the HIGGS~\cite{baldiSearchingExoticParticles2014} dataset with 1 MB
file per function.
%
%
Lambda functions are configured with 1 GiB RAM, and we run the training for 10 epochs.
We use autoscaling for DynamoDB and direct communication over TCP in FMI.

\emph{Performance.}
Figure~\ref{img:smi_benchmark_lambdaml} presents the distribution of
communication times per epoch, i.e., how much time passed between functions
ready to exchange data until they accumulate the centroids of the current epoch,
averaged across ten epochs.
FMI significantly reduces both the median and maximum communication time
by up to 105 and 1224 times, respectively, when running with 64 functions.
We did not increase the number of functions beyond 64 for DynamoDB due to
the timeouts and outliers we observed for 64 functions.
In contrast, FMI scaled well with good performance and few outliers up to 256 functions.
The performance benefits of using FMI stem from replacing DynamoDB
with a direct communication channel, and replacing a sequential reduction algorithm
of LambdaML with a parallel collective operation in FMI.
A further cause is the \code{base64}
serialization of binary data that is needed in communication using DynamoDB --- FMI directly operates on binary data.
Furthermore, we avoid unnecessary buffer copies by using \code{numpy} arrays
on top of existing memory buffers from the C++ library.

\begin{figure}[t]
    \centering
     \includegraphics[width=1\linewidth]{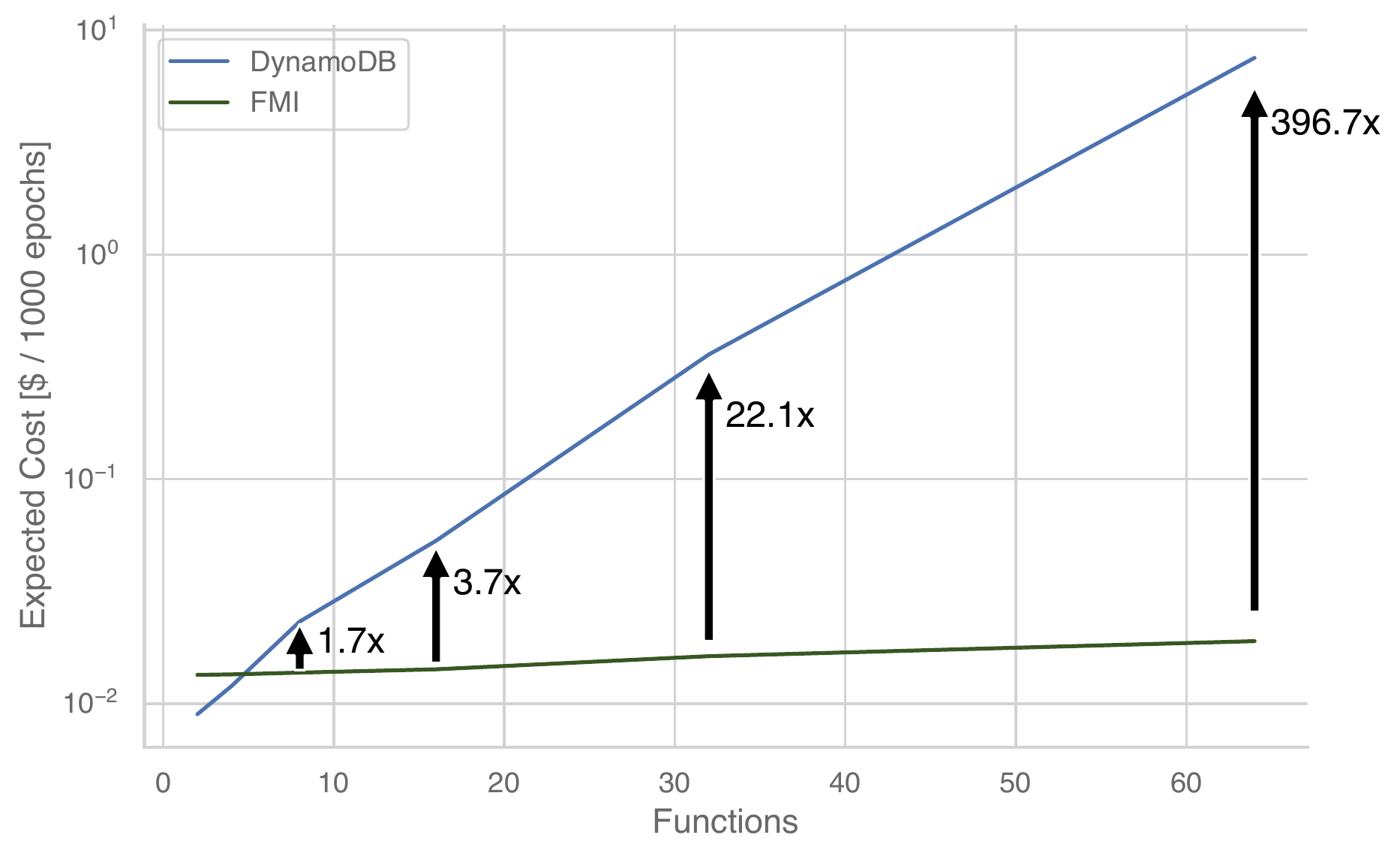}
    \caption{Cost comparison of FMI and the LambdaML DynamoDB channel. The annotation denotes the cost reduction provided by FMI.}
    \label{img:smi_benchmark_lambdaml_costs}
\end{figure}

\emph{Cost.}
We estimate the cost of moving data and running the Lambda function for the duration
of the communication epoch.
For FMI, we assume an hourly cost of running the hole punching service.
For DynamoDB, we assume one read and write unit per second for each function, as this is the minimum traffic AWS requires each function to provision.
This assumption is rather optimistic because one function can generate multiple
reads during one collective operation in LambdaML's implementation,
as functions repeatedly send requests until an item becomes visible.
%
%
Therefore, our estimation of the monetary benefits of using FMI is conservative.

Despite this pessimistic approximation, FMI results in significantly cheaper
communication costs (Fig.~\ref{img:smi_benchmark_lambdaml_costs}).
When using only a few functions, the cost is relatively similar because of the
the hourly cost of the hole punching server.
At 64 functions, a user pays approximately \$7.52 for 1000 epochs of
communication with DynamoDB while using FMI lowers these costs to less than \$0.02,
a reduction by a factor of 397.
This trend will increase with more functions as communication time increases significantly.
However, FMI infrastructure costs are practically independent of the number of functions,
due to the limited requirements of the hole punching server.
%

\emph{Integration.}
We integrate FMI into the K-Means benchmarks with \textbf{only four lines of code changed}.
%

%
\begin{bluebox}
  FMI can be integrated into serverless applications with minimal
  overhead, providing performance and cost improvements of two degrees of magnitude.
\end{bluebox}

\section{Related Work}

\label{sec:relatedwork}

Multiple works partially address communication in serverless environments and implement specialized
systems for given workloads (Table~\ref{tab:communication_solutions}).
In contrast, FMI provides a modular, high-performance, and general-purpose solution, with support for various communication channels and
a model-driven selection at runtime.

\textbf{Ephemeral Storage for Serverless.}
Pocket~\cite{10.5555/3291168.3291200} is a specialized data store for intermediate data in serverless,
with automatic resource scaling and multiple storage tiers.
Pocket is orthogonal to our work and can be integrated into FMI as a cheaper
alternative to in-memory stores.
Locus~\cite{pu2019shuffling} and Crucial~\cite{10.1145/3361525.3361535} include
specialized communication channels for serverless analytics and distributed
synchronization.
%
%

\begin{table}[t]
  \begin{adjustbox}{max width=\linewidth}
    \small
    \centering
    \begin{tabular}{lccccc}
      \toprule
      \multirow{2}{*}{Solution} & General & Object & In-Memory & Direct & Central \\
      & Purpose & Storage & Storage & Communication & Server \\
      \midrule
      Cirrus~\cite{10.1145/3357223.3362711} & & & & & \xmark{} \\
      Crucial~\cite{10.1145/3361525.3361535} & \xmark{} & & \xmark{} & & \\
      gg~\cite{234886} & \xmark{} & \xmark{} & \xmark{} & & \\
      Lambada~\cite{Mller2019LambadaID} & & \xmark{} & & & \\
      Boxer~\cite{wawrzoniak2021boxer} & \xmark{} & & & \xmark{} & \\
      LambdaML~\cite{jiang2021towards} & & \xmark{} & \xmark{} & & \xmark{} \\
      Locus~\cite{pu2019shuffling} & & & \xmark{} & & \\
      mu~\cite{10.5555/3154630.3154660} & \xmark{} & & & & \xmark{} \\
      Pocket~\cite{10.5555/3291168.3291200} & \xmark{} & & \xmark{} & & \\
      PyWren~\cite{DBLP:journals/corr/JonasVSR17} & \xmark{} & \xmark{} & & & \\
      Starling~\cite{perronStarlingScalableQuery2020} & & \xmark{} & & & \\
      \textbf{FMI (this work)}  & \xmark{}  & \xmark{}  & \xmark{}  & \xmark{}  & \xmark{}\\
      \bottomrule
    \end{tabular}
  \end{adjustbox}
    \caption{Comparison of serverless communication.}
    \label{tab:communication_solutions}
\end{table}

\textbf{Serverless Communication.}
Emerging frameworks support stateful and distributed FaaS jobs, but many of them focus
on domain-specific optimizations.
Systems such as gg~\cite{234886}, mu~\cite{10.5555/3154630.3154660},
and PyWren~\cite{DBLP:journals/corr/JonasVSR17}
are designed to handle general-purpose tasks, and they use cloud stores,
dedicated in-memory caches, and messaging servers.

%


Other systems target specific workloads and execution patterns.
LambdaML~\cite{jiang2021towards} and Cirrus~\cite{10.1145/3357223.3362711} are
specialized frameworks for machine learning, using custom parameter servers and dedicated stores for intermediate data.
Lambada~\cite{Mller2019LambadaID} and Starling~\cite{perronStarlingScalableQuery2020}
implements communication specialized for data analytics, including multi-level exchanges
and pipelining to minimize the cost and high latency of object storage operations.
Boxer~\cite{wawrzoniak2021boxer} extends Lambada with TCP hole punching.
While Boxer implements transparent hole punching for query processing, our solution offers
a collection of algorithms to target all serverless workloads that can benefit from inter-function communication.
Furthermore, we provide collective operations with an MPI-compatible interface to support HPC
applications.
Finally, the modular FMI system supports adding domain-specific communication optimizations,
similarly to the multitude of specializations for MPI collectives.

\textbf{Serverless Platforms.}
%
%
SONIC~\cite{mahgoub2021sonic} extends OpenLambda with application-aware data passing.
SAND~\cite{10.5555/3277355.3277444} implements a dedicated hierarchical message bus,
and Cloudburst~\cite{10.14778/3407790.3407836} adds co-located caches and an autoscaling key-value store.
%
%
The optimized communication channels available on a given platform can be integrated
into FMI, letting users benefit from the high performance of message-based
communication while hiding the complexity and specialization.


\section{Conclusions}

We propose FMI: an easy-to-use, modular, high-performance framework for general-purpose point-to-point and group communication in FaaS.
We benchmark communication channels available in serverless, implement direct TCP communication,
and derive performance and cost models that support the selection of optimal protocols.
FMI introduces collective communication to serverless, simplifying distributed computing
and allowing cloud providers to hide platform-specific optimizations. 
The FMI interface can be wrapped around existing MPI implementations,
improving the portability of applications between serverless functions and clusters.

We evaluate FMI by comparing the performance of its communication to MPI, and demonstrate the benefits of FMI in a case study of distributed machine learning,
showing easy integration and decreased communication time by up to 162x.
FMI brings serverless applications closer to the performance of MPI communication in HPC and lifts one of the most critical limitations of serverless computing.
%

\section*{Acknowledgments}

\setlength{\intextsep}{-5pt}
\setlength{\columnsep}{5pt}
\begin{wrapfigure}[2]{r}{.15\columnwidth} 
  \includegraphics[width=.12\columnwidth]{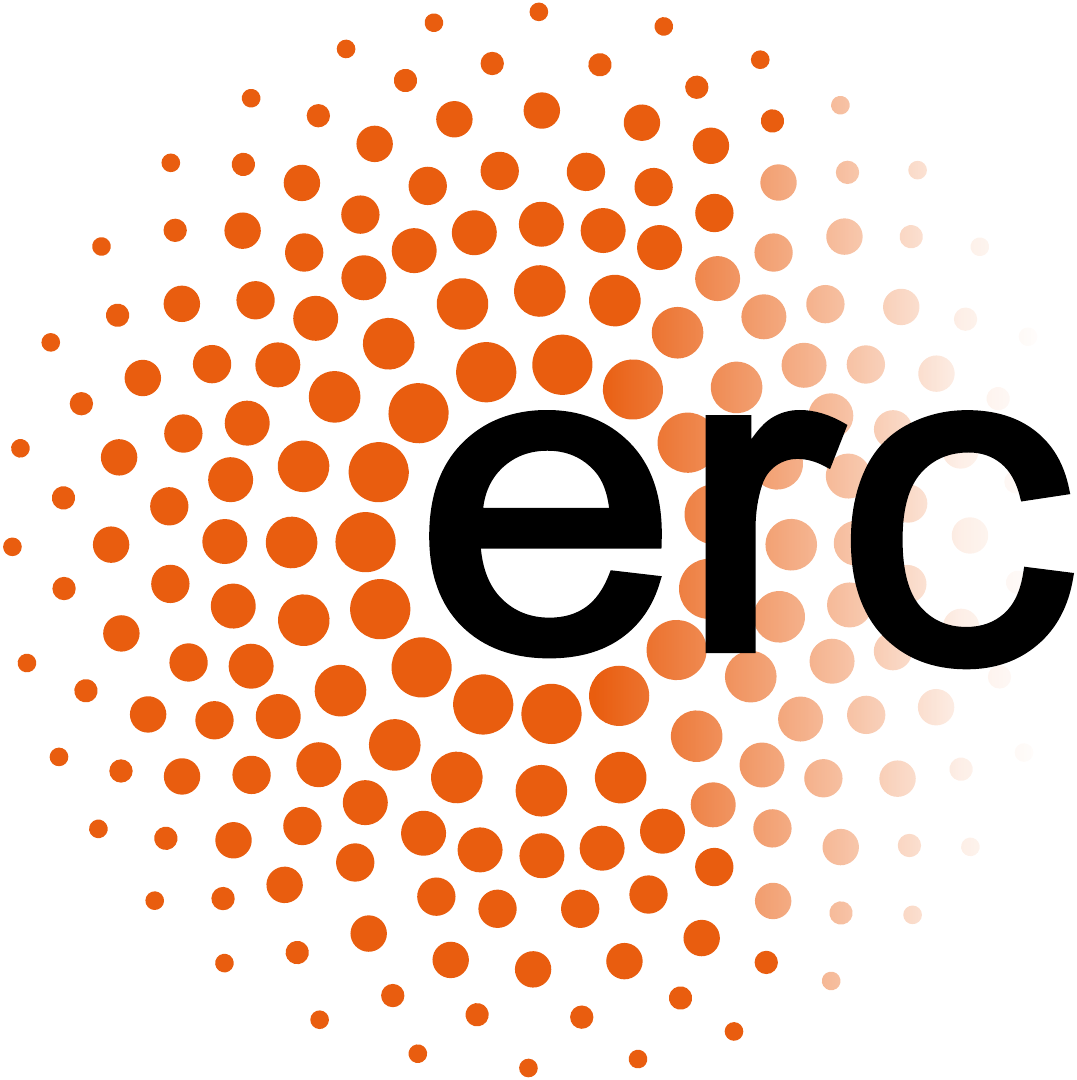}
\end{wrapfigure}
  %
  This project has received funding from the European Research Council (ERC)
  under the European Union's Horizon 2020 program (grant agreement PSAP, No. 101002047),
  and EuroHPC-JU funding under grant agreements DEEP-SEA, No. 95560 and RED-SEA, No. 955776).
  We thank Amazon Web Services for supporting this research with credits
  through the AWS Cloud Credit for Research program.

\bibliographystyle{ACM-Reference-Format}
\bibliography{serverless,cloud,mpi}

\end{document}